\documentclass[prb,twocolumn,aps]{revtex4}
\input{psfig}

\newcommand{\ket}[1]{\left|#1\right\rangle}

\begin{document}

\title{Spin-spin correlators in Majorana representation}

\author{Alexander Shnirman$^{1}$ and Yuriy Makhlin$^{1,2}$}

\affiliation{
$^1$Institut f\"ur Theoretische Festk\"orperphysik,
Universit\"at Karlsruhe, D-76128 Karlsruhe, Germany\\
$^2$Landau Institute for Theoretical Physics,
Kosygin st. 2, 117940 Moscow, Russia}

\begin{abstract}
In the Majorana representation of a spin 1/2 we find an identity which relates
spin-spin correlators to one-particle fermionic correlators. This should be
contrasted with the straightforward approach in which two-particle
(four-fermion) correlators need to be calculated. We discuss applications to the 
analysis of the dynamics of a spin coupled to a dissipative environment and of a 
quantum detector performing a continuous measurement of a qubit's state.
\end{abstract}
\maketitle

\section{Introduction}

An analysis of spin dynamics involves calculations of spin correlators.
Spin operators do not satisfy the Wick theorem, and various methods have been
used to still enable the use of perturbative (diagrammatic) methods. One of the
approaches is based on the Majorana-fermion representation of spin operators.
This approach has a long history and was applied recently to condensed-matter
problems.~\cite{Tsvelik_Majorana,Coleman_Miranda_Tsvelik_Majorana,Tsvelik_Book}

In this approach one introduces three Majorana fermions $\eta_{x,y,z}$ (per
spin) and expresses the spin (or Pauli) operators via these fermions:
\begin{eqnarray}
\label{Eq:Majorana_representation_3}
&&\sigma_{x} = -i\eta_y \eta_z\nonumber \\
&&\sigma_{y} = -i\eta_z \eta_x\nonumber \\
&&\sigma_{z} = -i\eta_x \eta_y
\ .
\end{eqnarray}
Obviously, an analysis of spin-spin correlations based on
Eq.~(\ref{Eq:Majorana_representation_3}) requires
calculations of a four-fermion correlator. In this note we show that spin
correlators coincide with certain two-fermion (i.e., one-particle) correlators.
We discuss under which conditions this identity may simplify the analysis and
discuss its particular applications.

\section{Majorana representation}

Majorana fermions satisfy
the anti-commutation relations: $\eta_\alpha\eta_\beta = - \eta_\beta
\eta_\alpha$ for $\alpha\ne\beta$ and $\eta_\alpha^2=1$, and are real:
$\eta_\alpha^\dagger = \eta_\alpha$ ($\alpha,\beta,\gamma = x,y,z$). These
properties ensure that the representation (\ref{Eq:Majorana_representation_3})
reproduces the commutation relations of the spin algebra.
An important feature of the representation (\ref{Eq:Majorana_representation_3}) 
is the fact that the spin operators
are bilinear (`bosonic') combinations of Dirac (annihilation / creation) 
fermionic operators.

One may construct the three Majorana fermion operators (for a spin) out of three 
different Dirac fermions, $\eta_\alpha = c^{\dag}_\alpha + c_\alpha$, each 
annihilation operator acting in its own two-dimensional Hilbert space.
This ensures the anticommutation relations and the property $\eta_\alpha^2 = 1$.
The whole Hilbert space is then 8-dimensional.

The dimensionality of the Hilbert space may be reduced if two Majorana fermions, 
e.g. $\eta_x$ and $\eta_y$, are constructed as linear combinations of a single 
Dirac fermion $f$ and its conjugate $f^\dagger$:
$\eta_x = f+f^{\dag}$ and $\eta_y=i(f^{\dag}-f)$.
Another Dirac fermion $g=c_z$ is
still needed to construct the third Majorana fermion $\eta_z = g+g^{\dag}$.
In this mixed Majorana-Dirac picture
\begin{eqnarray}
\label{Eq:Majorana_representation_2}
&&\sigma_{+} = \eta_z f\nonumber \\
&&\sigma_{-} = f^{\dag} \eta_z\nonumber \\
&&\sigma_{z} = 1-2f^{\dag}f
\ .
\end{eqnarray}
This `drone-fermion' representation was used for the analysis of magnetic
systems in the 60s
(see Refs.~\cite{Mattis,Kenan_Drone_Fermion,Spencer_Doniach,Spencer}).
The whole Hilbert space in this representation is 4-dimensional.
Depending on the rotational symmetry, this representation may be more convenient 
than (\ref{Eq:Majorana_representation_3}).

We have described this construction in the Hilbert space of two Dirac fermions 
$f$ and $g$. Alternatively one can view it as two
replicas of the original spin. Indeed, let us label the basis states in the 
following
way: denote the state without $f$- and $g$-fermions by $\ket{\uparrow_a}\equiv
\ket{00}$, and also $\ket{\uparrow_b}\equiv \ket{01}\equiv g^{\dag}\ket{00}$,
$\ket{\downarrow_b}\equiv \ket{10}\equiv f^{\dag}\ket{00}$, and
$\ket{\downarrow_a}\equiv \ket{11}\equiv f^{\dag}g^{\dag}\ket{00}$.
For the state $\ket{s_n}$ the first index $s=\uparrow/\downarrow$ denotes the 
spin component, while
the second $n=a/b$ labels the spin copy.
One notices that the spin operators $\sigma_+$, $\sigma_-$, and
$\sigma_z$ do not mix the $a$- and $b$-subspaces, i.e., they operate in the same 
way on the two `copies' of the spin. Further, one may view the index $n=a/b$ as 
an isospin and introduce the respective Pauli isospin operators $\tau_{x,y,z}$. 
In particular, $\tau_x$ is the `copy-switching' operator: $\tau_x\ket{s_a} = 
\ket{s_b}$ and $\tau_x\ket{s_b} = \ket{s_a}$.
The fermionic operators can be expressed as 
\begin{equation}
f=\sigma_+ \tau_x,\quad
f^{\dag}=\sigma_- \tau_x,\quad  \eta_z = \sigma_z \tau_x\,.
\end{equation}
Further, $\eta_\alpha = \sigma_\alpha\tau_x$, for any $\alpha=x,y,z$.
Accordingly,
in terms of the fermionic operators we have $\tau_x = (1-2f^{\dag}f)\,\eta_z= 
-i\eta_x\eta_y\eta_z$.
The operator $\tau_x$ commutes with `all other' operators: with
$\sigma_+$, $\sigma_-$, $\sigma_z$, $f$, $f^{\dag}$, and $\eta_z$.

\section{Reduction of the spin-spin correlators}

A physical Hamiltonian depends on the spin
operators $\sigma_x$, $\sigma_y$, $\sigma_z$ (and on 
other degrees of freedom, e.g., the electrons in the Kondo problem). Thus the 
operator $\tau_x$ commutes with any Hamiltonian and 
we obtain
\begin{eqnarray}
\label{Eq:Bubble_to_Line_Reduction}
&&\langle \sigma_\alpha(t)\sigma_\beta(t')\rangle
\nonumber\\
&&=\langle \tau_x(t)\eta_\alpha(t)\tau_x(t')\eta_\beta(t')
\rangle=
\langle\eta_\alpha(t)\eta_\beta(t')\rangle
\rangle\ .
\end{eqnarray}
Here we have used the fact that $\tau_x$ is time-independent, commutes with 
the Majorana fermions, and that $\tau_x^2=1$.
For example, we obtain 
\begin{eqnarray}
&&\langle \sigma_x(t)\sigma_x(t')\rangle=
\langle\eta_x(t)\eta_x(t')\rangle
\nonumber \\&&=
\langle [f(t)+f^{\dag}(t)]\,[f(t')+f^{\dag}(t')]\rangle
\ .
\end{eqnarray}

In certain situations the identity (\ref{Eq:Bubble_to_Line_Reduction}) may 
simplify the calculations. Indeed, if the Majorana representation 
(\ref{Eq:Majorana_representation_3}), (\ref{Eq:Majorana_representation_2}) is 
used for a calculation of spin-spin correlations, a four-fermion correlator 
needs to be evaluated. In a typical situation, the lowest-order contribution is 
given by a loop-like diagram which involves two fermionic propagators. A 
straightforward evaluation of higher-order contributions requires an analysis of 
the self-energy corrections to these propagators as well as of the vertex 
corrections. The use of the relation (\ref{Eq:Bubble_to_Line_Reduction}) reduces 
the task to the evaluation of a single self-energy.

In general, evaluation of this self-energy to all orders in the perturbative 
expansion involves complicated diagrams and, in particular, other self-energies 
and vertex corrections. Hence an involved calculation may still be needed. 
Nevertheless, the relation (\ref{Eq:Bubble_to_Line_Reduction}) may be useful if 
the needed self-energy part can be estimated reliably, for instance, by 
calculating the low-order contributions. We discuss two examples in 
Section~\ref{Sec:Example}.

Note also that the relation (\ref{Eq:Bubble_to_Line_Reduction}) may be 
straightforwardly generalized to multi-spin correlators, the evaluation of which 
then reduces to multi-fermion correlators. Further,
in a problem, that involves many spins represented via Majorana fermions, (e.g., 
a lattice spin model) the $\tau_x$-operators for different spins (sites) may be 
involved in a relation similar to (\ref{Eq:Bubble_to_Line_Reduction}). In this 
case their product does not drop out of the calculation (unlike $\tau_x^2=1$ for 
one spin).

\section{Gauge-invariance considerations}

One might be concerned by the fact that 
Eq.~(\ref{Eq:Bubble_to_Line_Reduction}) reduces
a correlator of physical quantities to 
a correlator of ``unphysical'' operators.
Another way of expressing this concern is to invoke 
the gauge symmetry. Indeed the Majorana 
representations (\ref{Eq:Majorana_representation_3}) and  
(\ref{Eq:Majorana_representation_2}) possess the discrete
$Z_2$-symmetry $\eta_\alpha \rightarrow -\eta_\alpha$.
As an operator which realizes the symmetry 
transformation one may choose $\tau_y = i(g^{\dag}-g)$
(cf.~Ref.~\cite{Coleman_Ioffe_Tsvelik_Majorana}). For 
example, $\eta_z \rightarrow \tau_y \eta_z \tau_y^{-1} = -\eta_z$.
Thus, the forth Majorana fermion allowed in the Hilbert space 
of the representation (\ref{Eq:Majorana_representation_2})
generates the symmetry transformation. One can as well use as a generator 
$\tau_z$, which just flips the sign of the wave function of the $b$-spin, 
keeping that of the $a$-spin intact.

Consider now a time-dependent
gauge transformation, which transforms a wave function $\ket{\Psi}$ to 
$U\ket{\Psi}$, 
where 
\begin{equation}
U=\exp\left(\frac{i\pi}{2}\,\tau_y\,\phi(t)\right)
\end{equation} 
and $\phi(t) = 0,1$. It transforms the operators
$\eta_\alpha \to (-1)^{\phi(t)} \eta_\alpha$,
$\tau_x \to (-1)^{\phi(t)} \tau_x$. Thus the operator 
$\tau_x$ is now time-dependent and no longer commutes 
with the Hamiltonian. Indeed, 
the gauge-transformed Hamiltonian reads
\begin{equation}
\tilde H = UHU^{-1} + i\dot U U^{-1} = H - \frac{\pi}{2}\,\dot \phi(t) 
\tau_y
\ ,
\end{equation}
and the last (gauge) term does not commute with $\tau_x$.
Thus we find that
\begin{equation}
\langle \eta_\alpha(t)\eta_\alpha(t') \rangle \to 
(-1)^{\phi(t)-\phi(t')} \langle \eta_\alpha(t) \eta_\alpha(t') \rangle
\ ,
\end{equation}
i.e., the single-fermion correlators are not gauge-invariant. 

The relation (\ref{Eq:Bubble_to_Line_Reduction})
is written in a fixed gauge, in which $\dot \phi(t) = 0$. In an arbitrary gauge, 
the factor $(-1)^{\phi(t)-\phi(t')}$ should be added on the rhs making the 
identity gauge-invariant.
In path-integral calculations
the integration is performed over all configurations (all gauges).
If a saddle-point solution breaks the $Z_2$-symmetry (and hence has a 
counterpart) fluctuations which involve switching between two mean-field 
solutions play an important part and need special 
attention.~\cite{Coleman_Ioffe_Tsvelik_Majorana}
However, for perturbative calculations around a trivial saddle point, e.g. above 
the Kondo temperature, the 
`fixed-gauge' relation (\ref{Eq:Bubble_to_Line_Reduction})
might be quite useful.

\section{Applications}
\label{Sec:Example}

The relation (\ref{Eq:Bubble_to_Line_Reduction})
can simplify calculations, since to evaluate a single-fermion 
Green function one needs only to evaluate a
self-energy. In a perturbative regime, when one may restrict oneself to 
evaluating the lowest-order contribution, 
Eq.~(\ref{Eq:Bubble_to_Line_Reduction}) simplifies the task.

One example of such a problem is dicussed in 
Ref.~\cite{Shnirman_Mozyrsky_MartinP} where continuous measurement of the state 
of a spin (qubit) by a quantum detector is analyzed and the ouput noise of the 
detector as well as the spin dynamics is studied. A calculation of a spin-spin 
correlator in the Majorana representation requires an evaluation of a 
two-fermion loop and involves the analysis of two self-energy parts and a vertex 
correction. It turns out that the vertex correction is important already in the 
lowest order and may even cancel one of the self-energy contributions. This fact 
may be understood and the calculations are considerably simplified if the 
relation~(\ref{Eq:Bubble_to_Line_Reduction}) is invoked. The obtained result 
coincides with an alternative calculation, in which instead of using the 
Majorana representation to enable the use of the Wick theorem one follows the 
dynamics of the spin directly. This approach~\cite{Schoeller_PRB} may be useful 
if the problem involves only a small number of spins (or other `non-Wick' 
degrees of freedom).

Here we consider another example: the dissipative dynamics of a spin coupled to 
an environment, for instance, the spin-boson model. Consider the case of a 
purely transverse coupling:
\begin{equation}
\label{Eq:Hspin-bath}
H = -\frac{1}{2}B\;\sigma_z
-\frac{1}{2}X(t)\;
\sigma_x +H_{\rm bath}
\ ,
\end{equation}
where $X$ is a fluctuating bosonic observable of the bath, whose  Hamiltonian 
$H_{\rm bath}$ determines the statistics of fluctuations.
We consider gaussian, but not necessarily equilibrium, fluctuations. 
Using the Keldysh technique we perform a calculation considering the bath-spin 
coupling (the second term on the rhs of Eq.~(\ref{Eq:Hspin-bath})) as a 
perturbation.

Below we use the notations of Ref.~\cite{Rammer_Smith}.
Let us introduce the matrix Green function of the bath:
\begin{equation}
\label{Eq:G_X}
G_{X} \equiv -i\langle T_{\rm K} X(t)X(t')\rangle
\end{equation}
Its Keldysh component
\begin{equation}
G_{X}^{K} = G_{X}^{>}+G_{X}^{<}=-2i S_X
\ ,
\end{equation}
and the difference between the retarded and advanced components
\begin{equation}
G_{X}^{R}-G_{X}^{A} = G_{X}^{>}-G_{X}^{<}=-2i A_X
\ ,
\end{equation}
are related to the symmetrized and antisymmetrized correlators: $\langle X(t) 
X(0) \rangle \equiv S_X(t) + A_X(t)$, $S_X(-t) = S_X(t)$, $A_X(-t) = - A_X(t)$. 
In equilibrium they are related by the fluctuation-dissipation theorem: 
$S_X(\omega) = A_X(\omega)\coth(\hbar \omega/2k_{\rm B}T)$. Here $G_{X}^{>} = 
-i\langle X(t)X(0) \rangle$ and
$G_{X}^{<} = -i\langle X(0) X(t) \rangle$.

Similarly, for the
Majorana fermion $\eta\equiv\eta_z$ we define $G_{\eta}\equiv -i\langle T_{\rm
K}\eta(t)\eta(t')\rangle$. The bare Green functions are $G_{\eta,0}^{>}
= -i$ and $G_{\eta,0}^{<} = i$. For the $f$-fermion we use the Bogolubov-Nambu 
spinors
$\Psi \equiv (f,f^{\dag})^{T}$ and $\Psi^{\dag}\equiv
(f^{\dag},f)$ and define a matrix $G_{\Psi}\equiv -i\langle T_{\rm
K}\Psi(t)\Psi^{\dag}(t')\rangle$.

Calculation of the spin propagators (in a stationary state) reduces due to the 
relation (\ref{Eq:Bubble_to_Line_Reduction}) to the evaluation of the
fermion Green functions $G_{\Psi}$ and $G_{\eta}$.
These functions can be found from the the Dyson (kinetic) equations:
\begin{equation}
\label{Eq:Dyson_Psi}
G_{\Psi}^{-1}=G_{\Psi,0}^{-1}-\Sigma_{\Psi}
\ ,
\end{equation}
\begin{equation}
\label{Eq:Dyson_eta}
G_{\eta}^{-1}=G_{\eta,0}^{-1}-\Sigma_{\eta}
\ ,
\end{equation}
where $\Sigma_{\Psi}$ and $\Sigma_{\eta}$ are the self-energies. 
All the quantities in Eq.~(\ref{Eq:Dyson_Psi})
are $4\times 4$ matrices (with the Nambu and Keldysh components).
In the frequency domain the operator $G_{\Psi,0}$ is given by
\begin{equation}
\label{Eq:G_Psi_0}
G_{\Psi,0}^{-1}=
\left(
\begin{array}{cccc}
\omega - B & 0 & 0 & 0\\
0 & \omega + B & 0 & 0\\
0 & 0 & \omega - B & 0\\
0 & 0 & 0 & \omega + B
\end{array}
\right)
\ ,
\end{equation}
while for $G_{\eta,0}$ we obtain
\begin{equation}
\label{Eq:G_eta_0}
G_{\eta,0}^{-1}=
\left(
\begin{array}{cc}
\omega/2 & 0 \\
0 & \omega/2 
\end{array}
\right)
\ .
\end{equation}
We disregard infinitesimal imaginary terms in 
Eqs.~(\ref{Eq:G_Psi_0}, \ref{Eq:G_eta_0})
since they are superseded by the self-energy parts in Eqs.~(\ref{Eq:Dyson_Psi}, 
\ref{Eq:Dyson_eta}).

\begin{figure}
\centerline{\hbox{\psfig{figure=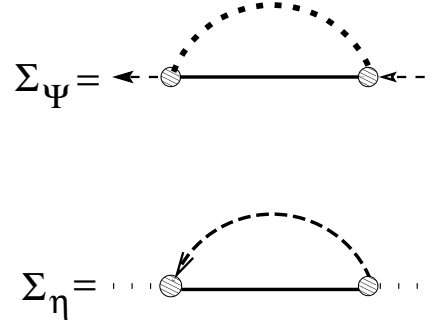,width=0.65\columnwidth}}}
\caption[]{\label{Figure:Sigma_ferm} The lowest order non-vanishing
contributions tor the self-energies $\Sigma_{\Psi}$ and $\Sigma_{\eta}$.
The solid lines denote the propagators of the bath ($X$), the dashed line ---
of the $f$-fermion, and the dotted lines --- of
the Majorana fermion $\eta$. (See text.)}
\end{figure}

The lowest-order contributions to the self-energies
\begin{equation}
\Sigma_{\Psi}=\left(\begin{array}{cc}
\Sigma_{\Psi}^{R} & \Sigma_{\Psi}^{K}\\
0 & \Sigma_{\Psi}^{A}
\end{array}\right)
\ ,\ \ \
\Sigma_{\eta}=\left(\begin{array}{cc}
\Sigma_{\eta}^{R} & \Sigma_{\eta}^{K}\\
0 & \Sigma_{\eta}^{A}
\end{array}\right)
\end{equation}
are shown in Fig.~\ref{Figure:Sigma_ferm} (the matrix structure after the 
Keldysh rotation is given~\cite{Rammer_Smith,Kamenev_Lectures}).
We find
that $\Sigma_{\Psi}^{>} = (i/4) \hat \lambda\,G_{X}^{>}\,G_{\eta}^{>}$
and $\Sigma_{\Psi}^{<} = (i/4) \hat \lambda\,G_{X}^{<}\, G_{\eta}^{<}$,
where
$\hat \lambda =
\left(
\begin{array}{cc}
\phantom{-}1 & -1\\
-1 & \phantom{-}1
\end{array}
\right)$ ia a matrix in the Nambu space. 
Similarly, we obtain $\Sigma_{\eta}^{>}=(i/4)
\,G_{X}^{>}\,\left(\begin{array}{cc}1&-1\end{array}\right)
G_{\Psi}^{>}\left(\begin{array}{c}\phantom{-}1 \\ -1\end{array}\right)
$ and $\Sigma_{\eta}^{<}=(i/4)
\,G_{X}^{<}\,\left(\begin{array}{cc}1&-1\end{array}\right)
G_{\Psi}^{<}\left(\begin{array}{c}\phantom{-}1 \\ -1\end{array}\right)$.
While the retarded and advanced components of the bare fermion Green functions
are known, its Keldysh component contains information about the distribution 
function and is sensitive to the environment (cf.~Ref.~\cite{Parcollet_Hooley}).

To perform the calculation, one introduces the distribution 
function $h_{\eta}$ via $G_{\eta}^{K}(\omega)=h_{\eta}(\omega)
\left(G_{\eta}^{R}(\omega)-G_{\eta}^{A}(\omega)
\right)$. Evaluating $\Sigma_{\Psi}$ from the first diagram in 
Fig.~\ref{Figure:Sigma_ferm}, we find
\begin{equation}
\Sigma_{\Psi}^{R}(\omega)-\Sigma_{\Psi}^{A}(\omega)=
\frac{\hat \lambda}{4}
\left[G_{X}^{K}(\omega) + h_{\eta}(0)(G_{X}^{R}(\omega)-G_{X}^{A}(\omega))
\right]
\ ,
\end{equation} 
\begin{equation}
\Sigma_{\Psi}^{K}(\omega)=
\frac{1}{4}\hat \lambda 
\left[(G_{X}^{R}(\omega)-G_{X}^{A}(\omega)) + h_{\eta}(0)G_{X}^{K}(\omega)
\right]
\ .
\end{equation}
The symmetry relation $h_{\eta}(-\omega)=-h_{\eta}(\omega)$ implies 
$h_{\eta}(0)=0$. Hence without evaluating $h_\eta$ we find
\begin{equation}
\Sigma_{\Psi}^{R}(\omega) - \Sigma_{\Psi}^{A}(\omega) = 
-\frac{i}{2}\,\hat\lambda\, S_X(\omega) 
\ ,
\end{equation}
and 
\begin{equation}
\Sigma_{\Psi}^{K}(\omega)=-\frac{i}{2}\,\hat\lambda\, A_X(\omega)
\ .
\end{equation}
The real parts of the retarded
and advanced self-energies give the non-equilibrium generalization
of the Lamb shift, i.e., renormalize the level splitting $B$.
This renormalization is small if the noise is weak and has a non-singular 
spectrum.

Substituting the self-energy $\Sigma_{\Psi}$ into the 
Dyson equation (\ref{Eq:Dyson_Psi}) we obtain
\begin{eqnarray}
\label{Eq:G_Psi_(-1)}
&&G_{\Psi}^{-1}=G_{\Psi,0}^{-1}\nonumber \\
&&-\left(
\begin{array}{cccc}
-i\Gamma(\omega) & i\Gamma(\omega) & -i A_X(\omega)/2 & i A_X(\omega)/2\\
i\Gamma(\omega) & -i\Gamma(\omega) &  i A_X(\omega)/2 &  -i A_X(\omega)/2\\
0 & 0 & i\Gamma(\omega) & -i\Gamma(\omega)\\
0 & 0 & -i\Gamma(\omega) & i\Gamma(\omega)
\end{array}
\right)\ ,
\nonumber \\
\end{eqnarray}
where $\Gamma(\omega)\equiv S_X(\omega)/4$.
Inverting Eq.~(\ref{Eq:G_Psi_(-1)}) we obtain
\begin{equation}
\label{Eq:G_Psi_RA}
G_{\Psi}^{R/A}=\frac{\left(
\begin{array}{cc}
\omega+B\pm i\Gamma(\omega) & \pm i\Gamma(\omega)\\
\pm i\Gamma(\omega) & \omega-B \pm i\Gamma(\omega)
\end{array}
\right)}
{(\omega^2-B^2) \pm 2i\omega\Gamma(\omega)}
\ ,
\end{equation}
\begin{equation}
\label{Eq:G_Psi_K}
G_{\Psi}^{K}=\frac{\frac{i A_X(\omega)}{2}\,\left(
\begin{array}{cc}
-(\omega+B)^2 & \omega^2-B^2\\
\omega^2-B^2 & -(\omega-B)^2
\end{array}
\right)}
{(\omega^2-B^2)^2 + 4\omega^2\Gamma(\omega)^2}
\ .
\end{equation}

From Eqs.~(\ref{Eq:G_Psi_RA}) and (\ref{Eq:G_Psi_K}) 
we see that, at least in this order of perturbation 
theory, we can use the relation 
$G_{\Psi}^{K}(\omega)=h_{\Psi}(\omega)
\left(G_{\Psi}^{R}(\omega)-G_{\Psi}^{A}(\omega)
\right)$, where $h_{\Psi}(-\omega)=-h_{\Psi}(\omega)$
and (for any pair $ij$ of Nambu indices)
\begin{equation}
h_{\Psi}(\omega)=\frac{\Sigma_{\Psi,ij}^{K}(\omega)}
{\Sigma_{\Psi,ij}^{R}(\omega)-\Sigma_{\Psi,ij}^{A}(\omega)}=
\frac{A_X(\omega)}{S_X(\omega)}
\ ,
\end{equation}
a quantity of the zeroth order in the coupling constant.
    
Using these results we can calculate $\Sigma_{\eta}$:
\begin{eqnarray}
&&\Sigma_{\eta}^{R}(\omega)-\Sigma_{\eta}^{A}(\omega) =
\frac{1}{8}
[G_{X}^{K}(\omega-B)
\nonumber \\
&&+
h_{\Psi}(B)(G_{X}^{R}(\omega-B)-G_{X}^{A}(\omega-B))
]
\nonumber \\
&&+[B\to -B]\ ,
\end{eqnarray}
\begin{eqnarray}
&&\Sigma_{\eta}^{K}(\omega)=
\frac{1}{8} 
\left[(G_{X}^{R}(\omega-B)-G_{X}^{A}(\omega-B))
\right. 
\nonumber\\
&& +\left. 
h_{\Psi}(B)G_{X}^{K}(\omega-B)
\right]
+[B\to -B]
\ .
\end{eqnarray} 
We evaluate the self-energy $\Sigma_\eta$ near the pole $\omega=0$ of the
Green function $G_{\eta}$: 
\begin{equation}
\Sigma_{\eta}^{K}(0)=0
\end{equation}
and
\begin{equation}
\Sigma_{\eta}^{R}(0)-\Sigma_{\eta}^{A}(0)=
-\frac{i}{2}\,S_X(B)\left[1-\frac{A_X^2(B)}{S_X^2(B)}\right]
\ .
\end{equation}

Now, with the acquired knowledge of single-fermion Green functions, we can 
evaluate various spin-spin correlators. We start with  
\begin{eqnarray}
\Pi_{xx} &\equiv& -i\langle T_{\rm K}\sigma_x(t)\sigma_x(t')\rangle
\rangle
\ ,
\end{eqnarray}
where the bosonic time-ordering is chosen for the spin components.
From Eq.~(\ref{Eq:Bubble_to_Line_Reduction})
we deduce that
\begin{equation}
\Pi_{xx}^{>}=
-i \langle \sigma_x(t) \sigma_x(t') \rangle =
\left(\begin{array}{cc}1&\ 1\end{array}\right)
G_{\Psi}^{>}\left(\begin{array}{c} 1 \\ 1\end{array}\right)
\ .
\end{equation}
Since $G^>_\Psi = \frac{1}{2} [G_\Psi^K+G_\Psi^R-G_\Psi^A]$, we obtain
\begin{equation}
\label{Eq:Pi_>}
i \Pi_{xx}^{>} =
\frac{(S_X(\omega)+A_X(\omega)) B^2}
{(\omega^2-B^2)^2 + 4\omega^2\Gamma(\omega)^2}
\ .
\end{equation}
This result describes the peaks in the spectrum of spin corelations at $\omega = 
\pm B$, with the width corresponding to the dephasing 
rate $T_2^{-1}=\Gamma(B)=S_X(B)/4$, thus
reproducing the form known, e.g., from the analysis of the Bloch-Redfield 
equations.~\cite{Nobel}

Similarly, for the Green function of the Majorana
fermion $\eta$ we obtain
\begin{equation}
G_{\eta}^{R/A}(\omega)=\frac{2}{\omega \pm 2i\tilde \Gamma}
\ ,
\end{equation}
where 
\begin{equation}
\tilde \Gamma \equiv \frac{S_X(B)}{4}\,\left[1-\frac{A_X^2(B)}{S_X^2(B)}\right]
\ .
\end{equation}
The Keldysh component vanishes, $G_{\eta}^{K}=0$, 
and we find
\begin{equation}
\label{Eq:Pi_zz}
i \Pi_{zz}^{>} \equiv \langle \sigma_z(t) \sigma_z(t')\rangle_\omega=
i G_{\eta}^{>}=\frac{4\tilde\Gamma}{\omega^2+4\tilde\Gamma^2}
\ .
\end{equation}
This may be compared to the peak shape obtained in the Bloch-Redfield 
approach.~\cite{Nobel} It can be expressed in terms of the average value 
$\langle \sigma_z \rangle = A_X(B)/S_X(B)$ of the $z$-component:
\begin{equation}
\label{Eq:Pi_zz_exact}
i\Pi_{zz}^{>}=\langle \sigma_z \rangle^2\,2\pi\delta(\omega)+
\left[1-\langle \sigma_z \rangle^2\right]\,\frac{2T_1^{-1}}{\omega^2 + T_1^{-2}}
\ ,
\end{equation}
where the relaxation time is given by $T_1^{-1} = 2 T_2^{-1}$.
The result (\ref{Eq:Pi_zz}) reproduces the high-frequency behavior of 
Eq.~(\ref{Eq:Pi_zz_exact}) and the `width of the peak', which may be read off 
from the high-$\omega$ asymptotics, $i\Pi_{zz} \sim 2 (1-\langle \sigma_z 
\rangle^2 )  / (T_1\omega^2)$. This indicates that at lower frequencies 
higher-order corrections (contributions to $\Sigma_\eta$) are important, which 
requires a further analysis.

\section{Summary}

In summary, we discussed an identity relating spin correlations to 
single-particle Majorana-fermion propagators and problems where the use of this 
relation simplifies the analysis.

We are grateful to  A.~Auerbach, T.~Costi, Y.~Gefen, M.~Kiselev, J.~Kroha, 
I.~Martin, and P.~W\"olfle for stimulating discussions, and in particular to
A.~Mirlin, A.~Rosch, and M.~Vojta for numerous inspiring conversations.
Y.M. was supported by the Humboldt foundation, the BMBF, and the ZIP Programme 
of the German government. 

{\bf Note added:} At the final stage of the preparation of this manuscript a 
preprint~\cite{Mao} appeared which partially overlaps with our work.

\end{document}